\documentstyle [12pt]{article}
\newcommand{\cs}[3]{{{#3} \brace {#1 #2}}}

\newcommand{\edf}{\ {\mathop{=}\limits^{\rm def}}\ }

\begin{document}

\begin{center}
 \bf{{Spinning and Spinning Deviation Equations for Special Types of Gauge Theories of Gravity}}
\end{center}
\begin{center}
 Magd E. Kahil\footnote{Modern Sciences and Arts University, Giza, Egypt \\
 Egyptian Relativity Group, Cairo, Egypt\\
  E-mail:mkahil@msa.eun.eg}
\end{center}
\begin{center}
\abstract{The problem of spinning and spin deviation equations for particles as defined by their microscopic effect has led many authors to revisit non-Riemannian  geometry  for being described torsion and its relation with the spin of elementary particles. We obtain a new method to detect the existence of torsion by deriving  the equations of spin deviations  in different classes of non-Riemannian geometries, using a modified Bazanski method.  We   find that translational gauge potentials and rotational gauge potentials regulate the spin deviation equation in the presence of Poincare gauge field theory of gravity.}
\end{center}
\section*{Introduction}
Einstein's legacy has bestowed geodesics and null geodesics  to examine the trajectory massive and massless particles respectively, introducing the notation of a test particle  which ignores the interaction associated with its intrinsic properties. Such a problem may be assigned to measure the behavior of a certain gravitational field. Yet, the concept of test particle is counted to  be existed relatively rather than absolutely. From this principle, the problem of spinning objects  is necessary to be examined using the Mathissson-Pappetrou equation [1].

Due to extending the geometry to become a non-Riemanian, torsion is expressed in some theories of gravity to be interacting with the spin of elementary particle, this is vital to examine the internal symmetry of some gauge theories of gravity with the flavor of Yang-Mills for this issue, one  introduces its own corresponding building blocks which mainly related to the tetrad space , in order to relate it with some properties of elementary particles [3-5]. Such theories   are developed in different stages from Utyima (1956), Kibble (1961), Sciama (1962),  Hehl et al (1976) [3-6]and finally crowned with MAG in 1995.[7] \\ The main theme of  these theories is centered on its  description  in the presence of the tetrad field  following same mechanism of Yang-Mills gauge theory  of spaces admitting non vanishing curvature and torsion  represented as  gauge theories of gravity [8].

 This approach has led  many authors to consider a wide spectrum of theories of gravity possessing gauge formulation such as Teleparallel-gravity [9], gauge version of GR in  tetrad space (torsion-less)[10], and most general one is the Poincare gauge field theory of gravity [11]. \\

\section{The Papapertrou Equation in General Relativity: Lagrangian formalism}
  The Lagrangian formalism of a spinning and precessing object and their corresponding deviation equation in Riemanian geometry is derived by the following Lagrangian [18]
\begin{equation}
 L= g_{\alpha \beta} P^{\alpha} \frac{D \Psi^{\beta}}{Ds} + S_{\alpha \beta}\ \frac{D \Psi^{\alpha \beta}}{Ds}+ F_{\alpha}\Psi^{\alpha}+ M_{\alpha \beta}\Psi^{\alpha \beta}
  \end{equation}
 where
 $$ P^{\alpha}= m U^{\alpha}+ U_{\beta} \frac{D S^{\alpha \beta}}{DS}.$$
 Taking the variation with respect to $ \Psi^{\mu}$ and $\Psi^{\mu \nu}$ simultaneously we obtain
 \begin{equation}
\frac{DP^{\mu}}{DS}= F^{\mu},
 \end{equation}
 \begin{equation}
\frac{DS^{\mu \nu}}{DS}= M^{\mu \nu} ,
 \end{equation}
 where $P^{\mu}$ is the momentum vector, $ F^{\mu} \edf \frac{1}{2} R^{\mu}_{\nu \rho \delta} S^{\rho \delta} U^{\nu},$ and $R^{\alpha}_{\beta \rho \sigma}$ is the Riemann curvature, $\frac{D}{Ds}$ is the covariant derivative with respect  to a parameter $S$,$S^{\alpha \beta}$ is the spin tensor, $ M^{\mu \nu} =P^{\mu}U^{\nu}- P^{\nu}U^{\mu}$, and $U^{\alpha}= \frac{d x^{\alpha}}{ds}$ is the unit tangent vector to the geodesic. \\
 Using the following identity on both equations (1) and (2)
  \begin{equation}
  A^{\mu}_{; \nu \rho} - A^{\mu}_{; \rho \nu} = R^{\mu}_{\beta \nu \rho} A^{\beta},
  \end{equation}
  where $A^{\mu}$ is an arbitrary vector.
 Multiplying both sides with arbitrary vectors, $U^{\rho} \Psi^{\nu}$ as well as using the following condition [Heydri-Fard et al (2005)]
 \begin{equation}
 U^{\alpha}_{; \rho} \Psi^{\rho} =  \Psi^{\alpha}_{; \rho } U^{\rho},
 \end{equation}
and $\Psi^{\alpha}$ is its deviation vector associated to the  unit vector tangent $U^{\alpha}$.
 Also in a similar way:

\begin{equation}
 S^{\alpha \beta}_{; \rho} \Psi^{\rho} =  \Phi^{\alpha \beta}_{; \rho } U^{\rho},
\end{equation}

 one obtains the corresponding deviation equations [19]
\begin{equation}
\frac{D^2 \Psi^{\mu}}{DS^2}= R^{\mu}_{\nu \rho \sigma}P^{\nu} U^{\rho} \Psi^{\sigma}+ F^{\mu}_{; \rho} \Psi^{\rho},
 \end{equation}
and
\begin{equation}
\frac{D\Psi^{\mu \nu}}{DS}=  S^{\rho [ \mu} R^{\nu ]}_{\rho \sigma \epsilon} U^{\sigma} \Psi^{\epsilon} + M^{\mu \nu}_{; \rho} \Psi^{\rho}.
 \end{equation}
Equations (1.7), (1.8) are essentially vital  to solve  the problem of stability for different celestial objects in various gravitational fields. This will examined in detail in our future work.

  \section{The Papapertrou Equation in Rieman-Cartan Theory of Gravity: Lagrangian formalism}
  The Mathisson-Papapetrou equation in non-Riemanian geometry are generalized forms of both (1.7)and(1.8), as a result of existence of a torsion tensor  $\Lambda^{\alpha}_{\beta \gamma}$, which is considered as a propagating field defined in the following manner,
  \begin{equation}
\Lambda^{\alpha}_{\beta \gamma} \edf \frac{1}{2} ( \delta^{\alpha}_{\beta}  \phi_{, \gamma} - \delta^{\alpha}_{\gamma}  \phi_{, \beta} ) \end{equation}
where $\phi$ is a scalar quantity.\\
 Yet, there are two different visions of admitting torsion in path  and spinning equations, one is considering it acting analogously as a Lorentz force, which led some authors to utilize the concept of torsion force [20]. Others may involve torsion in the affine connection by replacing the Christoffel symbol with the non-symmetric affine connection [21].
\subsection{Path and Path deviation equations having Torsion Force}
We suggest the following modified Bazanski Lagrangian to obtain the path and path deviation equations for non-Riemannian geometry using the notion of torsion force
 \begin{equation}
 L= g_{\alpha \beta} U^{\alpha}\frac{D \Psi^{\beta}}{DS} + \Lambda_{\alpha \beta \gamma} U^{\alpha} U^{\beta}\Psi^{\gamma}.
 \end{equation}
 Thus by taking the variation with respect to $\Psi^{\mu}$,
 provided that
$$
g_{\mu \nu ; \rho} =0,
$$
  we obtain,
 \begin{equation}
 \frac{d U^{\mu}}{ds} + \cs{\mu}{\nu}{\alpha} U^{\mu} U^{\nu} = - \Lambda_{\alpha \beta.}^{~.~.\mu} U^{\alpha} U^{\beta}
 \end{equation}

Using the following commutation relation
\begin{equation}
A_{\mu ; \nu \rho} - A_{\mu ; \rho \nu} = {R}^{\alpha}_{\beta \rho \sigma}A_{\alpha}
\end{equation}
on equations (2.11) provided that
\begin{equation}
\frac{D U^{\alpha}}{d \tau}= \frac{D \Psi^{\alpha}}{D S},
\end{equation}
we obtain the corresponding deviation equations
\begin{equation}
\frac{\nabla^{2}\Psi^{\mu}}{\nabla s^{2}} = R^{\mu}_{. \alpha \beta \gamma} U^{\alpha} U^{\beta} \Psi^{\gamma} + ( \Lambda_{\alpha \beta .}^{~.~.\mu} )_{;\rho}\Psi^{\rho}.
\end{equation}
\subsection{ Path and Path Deviation having torsion in a non-symmetric affine connection}
Path equations and path deviation equations are obtained its corresponding Bazanski approach as follows
\begin{equation}
L \edf g_{\mu \nu} \hat{U}^{\mu} \frac{\nabla \hat{\Psi^{\nu}}}{\nabla S},
\end{equation}
where,
\begin{equation}
\frac{\nabla \hat{\Psi}^{\mu \nu}}{\nabla S} = \frac{ d \hat{\Psi}^{\mu \nu}}{ d S} + \Gamma^{\alpha}_{. \beta \sigma} \hat{\Psi}^{\sigma},
\end{equation}
where
\begin{equation}
\Gamma^{\alpha}_{. \beta \sigma} \edf \cs{\beta}{\sigma}{\alpha} + K^{\alpha}_{.\beta \sigma}.
\end{equation}

By taking the variation with respect to $\Psi^{\alpha}$, provided that
\begin{equation}
g_{\mu \nu |\rho} =0,
\end{equation}
we obtain
\begin{equation}
\frac{\nabla \hat{U^{\alpha}}}{\nabla S} =   0,
\end{equation}

i.e.
$$
 \frac{d U^{\mu}}{ds} + \cs{\mu}{\nu}{\alpha} U^{\mu} U^{\nu} = - \Lambda_{\alpha \beta.}^{~.~.\mu} U^{\alpha} U^{\beta}.
 $$

Using the following commutation relation
\begin{equation}
A_{\mu || \nu \rho} - A_{\mu || \rho \nu} = \hat{R}^{\alpha}_{\beta \rho \sigma}A_{\alpha} + \Lambda^{\delta}_{\nu \rho}A_{\mu || \delta},
\end{equation}
and
\begin{equation}
\frac{\nabla \Psi^{\mu}}{\nabla \tau} =  \frac{\nabla U^{\mu}}{\nabla S }
\end{equation}
on equation (2.17)  we obtain
\begin{equation}
\frac{\nabla^2 \hat{\Psi^{\alpha}}}{\nabla S^2} = \hat{R}^{\alpha}_{\nu \rho \sigma} \hat{U}^{\nu} \hat{\Psi}^{\sigma} \hat{U}^{\rho} + \Lambda^{\rho}_{\mu \nu} U^{\alpha}_{| \rho} \hat{U}^{\mu}{\hat{\Psi}^{\nu}}.
\end{equation}

Thus, we found that equations (2.11) and (2.17) describe the same path equation but their corresponding  path deviation equations (2.13) and (2.20) are different due to the building blocks of the each type of geometry.

\subsection{Spin and spin deviation equations  having a torsion force}
For a spinning object, we suggest the following modified Bazanski Lagrangian, to derive both spin and spin deviation equations simultaneously.
\begin{equation}
L = g_{\mu \nu} P^{\mu} \frac{D \Psi^{\nu}}{DS}+ \Lambda_{( \mu \nu) \rho}P^{\mu} U^{\nu} \Psi^{\rho}+ S_{\mu \nu} \frac{D \Psi^{\mu \nu}}{DS} + \frac{1}{2} R_ {\mu \nu \rho \sigma}  S^{\rho \sigma}U^{\mu}\Psi^{\nu} + (P_{\mu}U_{\nu}- P_{\nu}U_{\mu})\Psi^{\mu \nu} .
\end{equation}
Taking the variation with respect to $\Psi^{\alpha}$ and $\Psi^{\alpha \beta}$ respectively we obtain
\begin{equation}
\frac{D P^{\alpha}}{DS} = - \Lambda_{(\mu \nu)}^{~..~ \alpha}P^{\mu}U^{\nu} + \frac{1}{2} {R}^{\alpha}_{\rho \mu \nu}S^{\mu \nu}U^{\rho},
\end{equation}
and
\begin{equation}
\frac{D S^{\alpha \beta}}{DS} = (P^{\alpha}U^{\beta} - P^{\beta}U^{\alpha}).
\end{equation}
The associated deviation equations are obtained by considering the following commutation relation
\begin{equation}
\frac{D}{DS} \frac{D}{D \tau} A^{\alpha} - \frac{D}{D \tau } \frac{D}{D S} A^{\alpha} = R^{\alpha}_{\beta \rho \sigma} A^{\beta} U^{\rho} \Psi^{\sigma}
\end{equation}
and
\begin{equation}
\frac{D U^{\alpha}}{DS}= \frac{D \Psi^{\alpha}}{D \tau},
\end{equation}
we get
\begin{equation}
\frac{D^2 \Psi^{\alpha}}{DS^2}  =  R^{\alpha}_{\beta \rho \sigma} U^{\beta} U^{\rho} \Psi^{\sigma}+ {( \frac{1}{2} R^{\alpha}_{\beta \mu \nu} S^{\mu \nu} U^{\beta} - \Lambda^{.~.~\alpha}_{. \mu\nu} )}_{\rho} \Psi^{\rho}
\end{equation}
and

\begin{equation}
\frac{D \Psi^{\alpha \beta}}{ D S} =  {S}^{\rho [ \beta }{R}^{\alpha ]}_{\rho \gamma \delta} U^{\gamma} \Psi^{\delta}_{; \delta} U^{\mu} \Psi^{\nu}  + ({P}^{\alpha}U^{\beta} - {P}^{\beta}U^{\alpha})_{; \rho} \Psi^{\rho}.
\end{equation}

\subsection{ Spin and spin deviation equations having torsion in a non-symmetric affine connection }

 If we replace the covariant derivative in Riemanian geometry by the absolute derivative in Einstein-Cartan geometry, we suggest the following Lagrangian of spinning and spinning deviation objects with precession
\begin{equation}
L = g_{\mu \nu} \hat{P}^{\mu} \frac{\nabla \hat{\Psi^{\nu}}}{\nabla S}+ S_{\mu \nu} \frac{\nabla \hat{\Psi^{\mu \nu}}}{\nabla S} + \frac{1}{2}\hat{R}_{\mu \nu \rho \sigma }\hat{S}^{\rho \sigma}U^{\nu}\Psi^{\mu} + (\hat{P_{\mu}}U_{\nu}- \hat{P_{\nu}}U_{\mu})\Psi^{\mu \nu},
\end{equation}
such that
\begin{equation}
\hat{P}^{\mu} \edf m \hat{U}^{\alpha} + \hat{U}_{\beta}\frac{\nabla S^{\alpha \beta}}{\nabla S},
\end{equation}
regarding that,
\begin{equation}
   \hat{S}^{\mu \nu} \hat{U}_{\nu} =0 \end{equation},

then taking the variation with respect to $\hat{\Psi^{\alpha}}$ and $\hat{\Psi^{\alpha \beta}}$. \\
Thus, we obtain
\begin{equation}
\frac{\nabla \hat{P^{\alpha}}}{\nabla S} =   \frac{1}{2}\hat{R}^{\alpha}_{\nu \rho \sigma} \hat{S}^{\rho \sigma} U^{\nu},
\end{equation}
and
\begin{equation}
\frac{\nabla \hat{S}^{\alpha \beta}}{\nabla S} = (\hat{P}^{\alpha}U^{\beta} - \hat{P}^{\beta}U^{\alpha}).
\end{equation}
Using  the commutation relation (2.22) and equation (2.27) on (2.33) and (2.34), we obtain the following set of deviation equations viz,

\begin{equation}
\frac{\nabla^2 \Psi^{\alpha}}{\nabla S^2} =  \hat{R}^{\alpha}_{\beta \gamma \delta}P^{\beta}U^{\gamma} \Psi^{\delta}+ { \Lambda^{\rho}_{\sigma \delta} U^{\sigma}\Psi^{\delta}P^{\sigma}}_{| \rho} +  \frac{1}{2} {( \hat{R}^{\alpha}_{\rho \sigma \epsilon} S^{\sigma \epsilon} U^{\rho})}_{| \rho} \Psi^{\rho},
\end{equation}
and

\begin{equation}
\frac{\nabla \Psi^{\alpha \beta}}{ \nabla S} =  \hat{S}^{\rho [ \beta }\hat{R}^{\alpha ] }_{\rho \gamma \delta} U^{\gamma} \Psi^{\delta} +  \Lambda^{\delta}_{\mu \nu}S^{\alpha \beta}_{| \delta} U^{\mu} \Psi^{\nu}  + (\hat{P}^{\alpha}U^{\beta} - \hat{P}^{\beta}U^{\alpha})_{| \rho} \Psi^{\rho}.
\end{equation}
We can find that there is a link between spin deviation tensor and torsion of space time. While, the spinning motion has no explicit relation with the same torsion tensor which confirms Hehl's point of view [22]. This result has led to find out its detailed description in case of taking into account the microstructure of any system. Such a trend can reach to define its contents with in the context of tetrad space.\\

\section{Tetrad Spaces and Gauge Theories of Gravity}

The concept of torsion of space time may give rise to  revisit its existence in a tetrad space, which may give rise to express  space-time as a system of two different coordinate systems. At each point of space-time is defined by the vector $x^{\mu}$ , $\mu =0,1,2,3$ and its metric tensor is $g_{\mu \nu}$. each point is associated with tangent space becoming a fiber of its corresponding tangent bundle given by Minkowski space whose metric tensor is defined by $\eta_{ab} \edf dig(1,-1,-1, -1)$ .

 Accordingly, this type of description may be analogous to explain the underlying geometry associated with some gauge theories e.g. it  is analogous to internal gauge theories, in which gravity becomes as a special gauge theory [12].  Thus, as a result of similarity between the above mentioned space-time and gauge theory, it is of interest to derive the gauge approach of equations of motion for different particles.\\
From this perspective, the problem of invariance of any quantities  must be a covariant derivative invariant under general coordinate transformation (GCT) and Local Lorentz transformation (LLT) that are expressible in terms of gauge potential of translation and rotation in the following way . \\
The building blocks of the space is two quantities, one represents the tetrad vector $(e^{a}_{\mu}$, and the other is the generalized spin connection $\Omega^{ij}_{\mu})$. The tangent space is raised and lowered by the Minkowski space,  while the space-time indices are raised and lowered by  the Riemannain metric

\begin{equation}
 g_{\mu \nu}  \edf e^{a}_{\mu}e^{b}_{\nu} \eta_{a b},
 \end{equation}

\begin{equation}
 g_{\mu \nu}  \edf e^{a}_{\mu}e^{b}_{\nu} \eta_{a b}.
 \end{equation}
This type of geometry defined its curvature tensor $R^{i}_{j \mu \nu}$ [25] is defined  as follows
\begin{equation}
R^{i}_{j \mu \nu} \edf \Omega^{i}_{j \nu ,  \mu}-\Omega^{i}_{j \mu ,  \nu} + \Omega^{k}_{j \nu \mu}\Omega^{i}_{k \mu} -\Omega^{k}_{j \mu \mu}\Omega^{i}_{k \nu}
\end{equation}

and its corresponding torsion tensor  $ \Lambda^{a}_{\mu \nu} $ is becoming as
\begin{equation}
\Lambda^{a}_{\mu \nu} \edf (e^{a}_{\mu, \nu}-e^{a}_{\nu, \mu} +  \Omega^{a}_{b \mu}e^{b}_{\nu} - \Omega^{a}_{b \nu}e^{b}_{\mu} )
\end{equation},
provided that
\begin{equation}
\nabla_{\nu}e^{m}_{\mu} =0,
\end{equation}
 i.e.
 $$
 e^{m}_{\mu || \nu } \equiv 0.
 $$
 Due to the following condition [23],
$$
g_{\mu \nu || \sigma} =0,
$$
which becomes
$$
e^{a}_{\mu, \nu} - \Gamma^{\lambda}_{\mu \nu} e^{a}_{\lambda} + \Omega^{a}_{.b \nu}e^{b}_{\mu} =0,
$$
and
  $$
    \Gamma^{\alpha}_{\nu \mu} e^{m}_{\alpha}  = e^{\alpha}_{m}( e^{m}_{\nu , \mu}   + \Omega_{\mu ~.~ n}^{~. m ~.}e^{n}_{\nu}).
 $$

Consequently, the relationship between  the generalized spin connection and contortion of space time can be obtained as\\
$$
 \Omega_{n.~ ~ \mu }^{~.m ~.} =  - e^{\nu}_{n}(e^{m}_{\nu , \mu}   +\Gamma^{m}_{\nu \mu})
$$
comparing with (2.17) one obtains that, one gets
$$
\Omega_{n.~ ~ \mu }^{~.m ~.}= \omega_{n.~ ~ \mu }^{~.m ~.} + K_{n.~ ~ \mu }^{~.m ~.}
$$

where $\omega_{n.~ ~ \mu }^{~.m ~.}$ spin connection associated  with Christoffel symbol, and
$$ K_{n.~ ~ \mu }^{~.m ~.} = e^{\nu}_{n} e^{m}_{\alpha}K^{\alpha}_{\nu \mu} . $$
 {\bf{Special cases:}} \\
 {i} Tele-parallelism: $\Omega^{i}_{j \mu} =0  \rightarrow  \omega^{i}_{j \mu} = - K^{i}_{j \mu}  $ gives that the spin connection is equivalent to Ricci coefficient of rotation\footnote{Ricci coefficient of rotation $A^{\rho}_{\mu \nu} \edf e^{\rho}_{i} e^{i}_{\mu ; \nu}$}. \\
 (ii) General Relativity in a gauge form: $K^{i}_{j \mu} =0  \rightarrow  \Omega^{i}_{j \mu} = \omega^{i}_{j \mu}  $ gives that the genealized spin connection is equivalent to Ricci coefficient of rotation.  \\

\subsection{ The Need for Gauge Theories in Gravity }
The study of microscopic structure of particles gives rise to utilize the richness Yang-Mills guage theory to express gravity as a gauge theory having internal gauge invariance.  This can be done by involving two different types of coordinate system one for GCT and the other for internal gauge invariance such as LLT. Consequently, it is worth mentioning that local Poincare gauge theory is an appropriate theory to explain gravity at this level. This approach was achieved by Hehl after a long process of versions by, Utimama, Kibble, Sciama, and others [11].\\
The advantage of importing Yang-Mills gauge field concept is the existence of  two field strength tensors  for translational and  rotational  in such a way that they are defined by two different vector potentials acting independently.
It is well known the analogy between gauge field and space time one can consider translational gauge $f^{a}_{\alpha}$  is equivalent to $e^{a}_{\mu}$ and the rotational gauge  $\Gamma^{ab}_{\beta}$ is equivalent to $\Omega^{a b}_{\mu}$. This may give rise to find another similarity between the gauge field strength $F^{ab}_{\mu \nu}$ and curvature of space time admitting the anholonomic coordinates $R^{a b}_{\mu \nu}$ [24].
to be added for Poincare gauge theory
Translational gauge potential $e^{n}_{\mu}$ and rotational gauge potential $\Gamma^{ab}_{\mu}$ in which the commutation derivative operators[11].
 As in gauge theories the commutation relation is defined as with respect to the gauge field strength

\begin{equation}
[ \tilde{D_{a}}, \tilde{D_{b}} ] = f^{\alpha}_{a}f^{\beta}_{b}( F^{.~.~. \mu \nu}_{\alpha \beta}s^{\alpha \beta} - F^{.~.~n}_{\alpha \beta}\tilde{D_{n}}),
\end{equation}
 which becomes is equivalent to
\begin{equation}
 [\nabla_{a},\nabla_{b}] = e^{\alpha}_{a}e^{\beta}_{b}( R^{.~.~. \mu \nu}_{\alpha \beta}s^{ab} - \Lambda^{.~.~n}_{\alpha \beta}\nabla_{n}).
\end{equation}

\subsection{Teleparalellism: Translational Gauge Theory}
In this case, $\Omega^{a}_{b \mu} =0$ the conventional absolute parallelism geometry: a pure gauge theory for translations [9],[12].\\
Using Acros and Pirra method[12], one can find out that

 $$
{\hat{R}}^{\alpha}_{\beta \gamma \delta} \edf \Gamma^{\alpha}_{\beta \gamma, \delta } - \Gamma^{\alpha}_{\beta \delta, \gamma } + \Gamma^{\epsilon}_{\beta \gamma} \Gamma^{\alpha}_{\epsilon \delta} - \Gamma^{\epsilon}_{\beta \delta} \Gamma^{\alpha}_{\epsilon \gamma} \equiv 0 ,
$$

$$  \Gamma^{\rho}_{\mu \nu}= e^{\rho}_{c}(e^{c}_{\mu , \nu} - e^{c}_{\nu , \mu}) , $$

$$\Lambda^{\alpha}_{\beta \gamma} = \Gamma^{\alpha}_{\beta \rho}- \Gamma^{\alpha}_{\rho \rho}. $$

As, in GR the spin connection is equal to Ricci coefficient of rotation $ \omega^{a}_{b \mu} \edf e^{a}_{\rho}e^{\rho}_{a ; \mu}   $
 thus,
 $$
 \omega^{a}_{b \mu} \edf A^{a}_{b \mu} - \gamma^{a}_{b \mu},
 $$
where $K^{a}_{b  \mu}$ its  contortion defined as
$$
K^{a}_{b  \mu} =\frac{1}{2}e^{c}_{\mu} (\Lambda^{~ a~}_{c . b } + \Lambda^{~a~}_{b . c} - \Lambda^{a~~}_{ . b c  }).
$$
 Consequently, one obtains,
$$
 R^{c}_{d \mu \nu} = \hat{R}^{c}_{d \mu \nu} - K^{c}_{d \mu \nu},
$$
where
 $$ K^{c}_{d \mu \nu} \edf  \gamma^{c}_{d \nu  ; \mu } - \gamma^{c}_{d \mu ; \nu} + \gamma^{c}_{a \mu} \gamma^{a}_{ d \nu} - \gamma^{c}_{a \nu} \gamma^{a}_{ d \mu}$$.

\subsection{General Relativity: A Tetrad Version of Gravitational Gauge Theory}
Collins et al (1989)[10] described GR as a gauge theory of gravity subject to the following gauge potential vectors $e^{a}_{\mu}$ and $\omega^{ij}_{.~. \mu}$ to represent translational and rotational gauge potentials respectively. \\
The equations of physics will contain derivatives of tensor fields and it is therefore necessary to define the covariant derivatives of tensor fields under the transformations GCT and LLT, one must need to define two types of connection fields to be associated with each of them. Accordingly the Christoffel symbol $\cs{\mu}{\nu}{\alpha}$ is referred to GCT while the spin connection $\omega_{a b \mu}$ as related to LLT. \\

which is considered a torsion less condition of  Poincare gauge theory

 \begin{equation}
 D_{\mu}e^{m}_{\nu}  \edf e^{m}_{\nu , \mu} - \cs{\nu}{\mu}{\alpha}e^{m}_{\alpha} + \omega_{\mu~.~n}^{~.m ~.}e^{n}_{\nu},
\end{equation}

 provided that
 \begin{equation}
  D_{\rho}D_{\mu}e^{m}_{\nu}=  D_{\mu}D_{\rho}e^{m}_{\nu}.
  \end{equation}

 Using this concept it turns out that GR may be expressed in terms of connecting $e^{\mu}_{a}$ , $\omega_{ a b \mu }$ and $\cs{\mu}{\nu}{\alpha}$ together
\begin{equation}
 g_{\mu \nu}  \edf e^{a}_{\mu}e^{b}_{\nu} \eta_{a b},
 \end{equation}
 such that,
\begin{equation}
 \cs{\mu}{\nu}{\alpha} \edf \frac{1}{2} g^{\alpha \sigma}(g_{\nu \sigma , \alpha}+g_{\sigma \alpha, \nu} -g_{\alpha \nu , \sigma}  )
 \end{equation}.
 Thus, the curvature tensor may be defined ,due to gauge approach,  in terms of spin connection $\omega^{a}_{b \mu}$
$$
 R^{c }_{.d \mu \nu} \edf  \omega^{c}_{d \nu , \mu} - \omega^{c}_{d \mu , \nu} + \omega^{c}_{a \mu }\omega^{a}_{d \nu} - \omega^{c}_{a \nu }\omega^{a}_{d \mu}.
$$
Using this concept it turns out that GR may be expressed in terms of connecting $e^{\mu}_{a}$ , $\omega_{ a b \mu }$ and $\cs{\mu}{\nu}{\alpha}$ together.
 Accordingly where $ R^{\alpha}_{.\mu d c}$ the curvature tensor may be defined ,due to gauge approach,  in terms of spin connection $\omega^{a}_{b \mu}$
$$
 R^{c }_{.d \mu \nu} \edf  \omega^{c}_{d \nu , \mu} - \omega^{c}_{d \mu , \nu} + \omega^{c}_{a \mu }\omega^{a}_{d \nu} - \omega^{c}_{a \nu }\omega^{a}_{d \mu}.
$$

\section{Motion of Special Classes for Gauge Theories}

\subsection{Path and Path Deviation Equations of Gauge Theories of Torsion Force}
We suggest the following Lagrangian to derive both path and path deviation equation for gauge theories having a torsion force.
\begin{equation}
L = e^{\mu}_{a} e^{\nu}_{b} U^{a} \frac{D \Psi^{b}}{DS}+ e^{\alpha}_{a} \Lambda_{\alpha \beta \gamma} P^{a}U^{\beta}\Psi^{\gamma},
\end{equation}

to obtain its corresponding path equations by applying the Bazanski approach to become,
\begin{equation}
\frac{D {U^{a}}}{D S} = - e^{a}_{\alpha}\Lambda^{~.~.~ \alpha}_{(\beta \gamma).} U^{\beta} U^{\gamma}.
\end{equation}

 Applying the following commutation relations of equation (2.20) and following (2.21), we obtain after some manipulation its corresponding path deviation equations

\begin{equation}
\frac{D \Phi^{a}}{DS} = e^{a}_{\mu} K^{\mu}_{\alpha \beta \sigma} U^{\alpha} U^{\beta} \Phi^{\sigma} +  \Lambda{.~.~a}_{.\beta \gamma} U^{\gamma}U^{\beta} )_{; \rho}\Phi^{\rho}.
\end{equation}

\subsection{Spin and Spin Deviation Equations of Gauge Theories of Torsion Force}
Thus spinning and spinning equations are obtained from taking the variation with respect to $\Psi^{\mu}$ and $\Psi^{\mu \nu}$  simultaneously, for the following Lagrangian
\begin{equation}
L = e^{\mu}_{a} e^{\nu}_{b} P^{a} \frac{D \Psi^{b}}{DS}+\Lambda_{a \beta \gamma} P^{a}U^{\beta}\Psi^{\gamma} - \frac{1}{2}  K_{ ab \mu  \nu  }  S^{ \mu \nu}U^{b} \Psi^{a} + S_{a b} \frac{D \Psi^{a b}}{DS} + (P_{a}U_{b}- P_{b}U_{a})\Psi^{ab}
\end{equation}

to obtain

\begin{equation}
\frac{D P^{a}}{DS} = - \frac{1}{2}  K^{a}_{  \nu \rho \delta } S^{\rho \delta}U^{\nu} + \Lambda{.~.~ a}_{.\beta \gamma} P^{\alpha}U^{\beta}.
\end{equation}
and

\begin{equation}
 \frac{D e^{a}_{\alpha} e^{b}_{\beta} S^{a b}}{DS} = ( P^{\alpha} U^{\beta}- P^{\beta}U^{\alpha}),
\end{equation}
and
\begin{equation}
 \frac{D  S^{c d}}{DS}= ( P^{c} U^{d}- P^{d}U^{c}).
\end{equation}
Using the commutation relations as mentioned above, we obtain;
\begin{equation}
\frac{D \Phi^{a}}{DS} = K^{a}_{b \beta \sigma} P^{b} U^{\beta} \Phi^{\sigma} + (R^{a}_{ b \mu \nu } S^{\mu \nu} U^{\alpha}+ \Lambda{.~.~a}_{.\beta \gamma} P^{\gamma}U^{\beta} )_{; \rho}\Phi^{\rho},
\end{equation}
and
\begin{equation}
\frac{D \Psi^{c d}}{DS} =  S^{\rho [ d }\hat{R}^{c ] }_{\rho \gamma \delta} U^{\gamma} \Psi^{\delta} + (P^{c}U^{d} - P^{d}U^{c})_{; \rho} \Psi^{\rho}.
\end{equation}

\subsection{Path and Path Deviation Equation of Gauge Theories of Torsion-less }
 The Lagrangian formalism of path and path deviation equation of gauge theories of torsion-less is given as
\begin{equation}
L \edf e^{\mu}_{a}e^{\nu}_{b} {U}^{a} \frac{D{\Psi^{b}}}{D S}.
\end{equation}

Applying the Bazanski approach , taking the variation with respect to $\Psi^{\alpha}$,
we obtain
\begin{equation}
\frac{D e^{\alpha}_{c}\hat{  U^{c}}}{D S} =   0,
\end{equation}
provided that
$$
\frac{D e^{a}_{\alpha} }{D S} \equiv 0,
$$
we get
\begin{equation}
\frac{D U^{c}}{D S} =   0.
\end{equation}

Applying the following commutation relation
\begin{equation}
A_{a || \nu \rho} - A_{ a || \rho \nu} \edf \hat{R}^{c}_{ b \rho \sigma}A_{b},
\end{equation}
on equation (4.59) and using (2.13) we obtain
\begin{equation}
\frac{D^2 \Psi^{c}}{D S^2} = {R}^{c}_{ b \rho \sigma} \hat{U}^{b} \hat{\Psi}^{\sigma} \hat{U}^{\rho}
\end{equation}
Substituting with the path equation in the deviation equation we get
\begin{equation}
\frac{D^2{\Psi^{c}}}{D S^2} = {R}^{c}_{b \rho \sigma} {U}^{b} \hat{\Psi}^{\sigma} {U}^{\rho}.
\end{equation}

\subsection{Spin and spin deviation equation of gauge theories of torsion-less }
Thus spinning and spinning equations are obtained from taking the variation with respect to $\Psi^{\mu}$ and $\Psi^{\mu \nu}$  simultaneously, for the following Lagrangian
\begin{equation}
L =e^{\mu}_{a}e^{\nu}_{b} P^{a} \frac{D \Psi^{b}}{DS}+\frac{1}{2} ( R_{ a b \mu \nu } ) S^{ \mu \nu}U^{b} \Psi^{a} + e^{a}_{\mu} e^{b}_{\nu} S_{a b} \frac{D \Psi^{\mu \nu}}{DS} + e^{\mu}_{a}e^{\nu}_{b}(P_{a}U_{b}- P_{a}U_{b})\Psi^{\mu \nu},
\end{equation}

to obtain its corresponding path equation using the Bazanski approach to get

\begin{equation}
\frac{D P^{a}}{DS} = \frac{1}{2} (R^{a}_ {  b \mu \nu}  )S^{\mu \nu}U^{\nu},
\end{equation}
and

\begin{equation}
 \frac{D  S^{a b}}{DS} = ( P^{\alpha} U^{\beta}- P^{\beta}U^{\alpha}),
\end{equation}

\begin{equation}
e^{a}_{\alpha} e^{b}_{\beta} \frac{D  S^{a b}}{DS}+ S^{ab}{} \frac{D ({e^{a}_{\alpha} e^{b}_{\beta}})}{DS}= ( P^{\alpha} U^{\beta}- P^{\beta}U^{\alpha}).
\end{equation}
Multiplying both sides by $ e^{c}_{\alpha} e^{d}_{\beta} $
\begin{equation}
e^{c}_{\alpha} e^{d}_{\beta} e^{a}_{\alpha} e^{b}_{\beta} \frac{D  S^{a b}}{DS}+ e^{c}_{\alpha} e^{d}_{\beta}S^{ab} \frac{D ({e^{a}_{\alpha} e^{b}_{\beta}})}{DS}= e^{c}_{\alpha} e^{d}_{\beta}( P^{\alpha} U^{\beta}- P^{\beta}U^{\alpha}),
\end{equation}
such that
$$
 \frac{D ({e^{a}_{\alpha} e^{b}_{\beta}})}{DS}=0.
$$
 Consequently, one obtains
\begin{equation}
 \frac{D  S^{c d}}{DS}= ( P^{c} U^{d}- P^{d}U^{c}).
\end{equation}
Accordingly, the deviation equation is obtained by applying the commutation relation and (2.27) on (4.52) and (4.54);
\begin{equation}
\frac{D \Phi^{\mu}}{DS} = R^{\mu}_{\alpha \beta \sigma} P^{\alpha} U^{\beta} \Phi^{\sigma} + (R^{\mu}_{\alpha a b} S^{a b} U^{\alpha}+ \Lambda^{.~.~ \alpha}_{.\beta \gamma} P^{\alpha}U^{\beta} )_{; \rho}\Phi^{\rho},
\end{equation}
and
\begin{equation}
\frac{D \Psi^{c d}}{DS} =  S^{\rho [ d }\hat{R}^{c ] }_{\rho \gamma \delta} U^{\gamma} \Psi^{\delta} + (P^{c}U^{d} - P^{d}U^{c})_{; \rho} \Psi^{\rho}.
\end{equation}

\subsection{ Path and Path Deviation Poincare Gauge Theory}
Path equations and path deviation equations are obtained its corresponding Bazanski approach as follows
\begin{equation}
L \edf e^{\mu}_{a}e^{\nu}_{b} \hat{U}^{a} \frac{\nabla \hat{\Psi^{b}}}{\nabla S}.
\end{equation}

By taking the variation with respect to $\Psi^{\alpha}$,
we obtain
\begin{equation}
\frac{\nabla e^{\alpha}_{c}\hat{  U^{c}}}{\nabla S} =   0,
\end{equation}
provided that
$$
\frac{\nabla e^{a}_{\alpha} }{\nabla S} \equiv 0,
$$
we get
\begin{equation}
\frac{\nabla \hat{U^{c}}}{\nabla S} =   0.
\end{equation}

Thus,  the spin deviation equations are obtained by applying the commutation relation and (2.17) on both
\begin{equation}
\frac{\nabla^2 \hat{\Psi^{c}}}{\nabla S^2} = \hat{R}^{c}_{ b \rho \sigma} \hat{U}^{b} \hat{\Psi}^{\sigma} \hat{U}^{\rho} +  e^{\delta}_{b} \Lambda^{b}_{\nu \rho}U_{ a || \delta}
\end{equation}
Substituting with the path equation in the deviation equation we get
\begin{equation}
\frac{\nabla^2 \hat{\Psi^{c}}}{\nabla S^2} = \hat{R}^{c}_{b \rho \sigma} \hat{U}^{b} \hat{\Psi}^{\sigma} \hat{U}^{\rho}
\end{equation}
\subsection{ Spin and Spin Deviation Equation Poincare Gauge Theory}

We suggest the following Lagrangian of spinning and spinning deviation objects with precession by replacing the covariant derivative in (1.1) by the absolute derivative as described in (2.15) to get
\begin{equation}
L = e^{\mu}_{a}e^{\nu}_{b}  \hat{P}^{a} \frac{\nabla \hat{\Psi^{b}}}{\nabla S}+ S_{ab} \frac{\nabla \hat{\Psi^{ab}}}{\nabla S} + \frac{1}{2}\hat{R}_{a b \mu \nu }\hat{S}^{ab}U^{\nu}\Psi^{\mu} + (\hat{P_{a}}U_{b}- \hat{P_{b}}U_{a})\Psi^{ab},
\end{equation}
such that
\begin{equation}
\hat{P}^{\mu} \edf m \hat{U}^{\alpha} + \hat{U}_{\beta}\frac{\nabla S^{\alpha \beta}}{\nabla S},
\end{equation}

 Taking the variation with respect to $\hat{\Psi^{\alpha}}$ and $\hat{\Psi^{\alpha \beta}}$ and after some manipulations,

 we obtain

\begin{equation}
\frac{\nabla \hat{P^{a}}}{\nabla S} =   \frac{1}{2}\hat{R}^{a}_{b  \rho \sigma} \hat{S}^{\rho \sigma} U^{b},
\end{equation}
and
\begin{equation}
\frac{\nabla \hat{S}^{a b}}{\nabla S} = (\hat{P}^{a}U^{b} - \hat{P}^{b}U^{a}).
\end{equation}

And their corresponding deviation equations are obtained by applying the commutation relation (2.20) and (2.21) on both (4.79) and (4.79) to get
\begin{equation}
\frac{\nabla^2 \Psi^{a}}{\nabla S^2} =  \hat{R}^{a}_{b \beta \gamma}P^{b}U^{\beta} \Psi^{
\gamma}+ { e^{\rho}_{c} \Lambda^{c}_{\beta \gamma} U^{\beta}\Psi^{\gamma}P^{a}}_{| \rho} +  \frac{1}{2} {( \hat{R}^{a}_{ b \rho \sigma } S^{\rho \sigma} U^{b})}_{| \delta} \Psi^{\delta},
\end{equation}
and

\begin{equation}
\frac{\nabla \Psi^{a b}}{ \nabla S} =  e^{b}_{\beta}e^{c}_{\delta}\hat{S}^{\delta [ \beta }\hat{R}^{ a ] }_{c \gamma \delta} U^{\gamma} \Psi^{\delta} + e^{\delta}_{c} \Lambda^{c}_{\mu \nu}S^{\alpha \beta}_{| \delta} U^{\mu} \Psi^{\nu}  + (\hat{P}^{a}U^{b} - \hat{P}^{b}U^{a})_{| \rho} \Psi^{\rho}.
\end{equation}
Thus, we see clearly the interaction between spin deviation tensor and torsion of space time is expressed in terms of curvature (rotational) and torsion (translational) strength fields.

\section{Discussion and Concluding Remarks}
In our present work, we have developed he Bazanski approach to obtain spin and spin deviation equations in non-Riemanian geometry. This approach was applied for obtaining path equations for some geometries admitting non vanishing curvature and torsion simultaneously. [26-28]\\
Due to the resultant equations (2.35),(2.36), we have figured out that torsion is explicitly mentioned in  spin deviation equations ,even if one puts $P^{\mu}= m U^{\mu}$,. Such a result is in favor of Hehl's argument falsifying the measurement of torsion from identifying the spin tensor, for a spinning object in an orbit. This result is an alternative approach to measure torsion from spinning equations using non-minimal coupling without introducing micro-structure[29].\\
Accordingly, we have obtained (2.35) and (2.36) of Poincare gauge theory of gravity as described within a  tetrad  space. Consequently, using the analogy between tetrads and gauge theories, we obtained equations  (4.82), (4.83) that show how a  tetrad vector $e^{\mu}_{i}$ is equivalent to gauge translational potential   explicitly mentioned in the equation while the generalized spin tensor $ \Omega^{ij}_{\mu}$, which is equivalent to gauge rotational potentials as mentioned implicitly  in the presence of curvature and torsion of space-time simultaneously. These equations are considered to be the generalized cases for some special classes of gauge theory of gravity of path and spin deviation equations  (4.55) and (4.56) having a torsion force,  as well as  their counterpart  of  (4.69) and (4.70) for a gauge theory of gravity having a torsion free.\\
Thus, we conclude that torsion of space time can be tested for any spinning object in any type of gravitational fields by examining its spin deviation equations.

\section*{Acknowledgement}
The author would like to thank Professor  T.Harko his remarks and comments.

\section*{References}
{[1]} A.Papapetrou , Proceedings of Royal Society London A {\bf{209}} , 248(1951).  \\
{[2]}E. Corinaldesi and A. Papapetrou  Proceedings of Royal Society London A {\bf{209}}, 259 (1951)\\
 {[3]}Utyiama, R.(1956) Phys Rev. 101 1597 \\
{[4]}Kibble, T.W. (1960) J. Math Phys., 2, 212 \\
{[5]}Hehl,F. W., von der Heyde, P. Kerlik, G.D. and Nester, J.M. (1976) Rev Mod Phys 48, 393-416 \\
{[6]} Hehl, F.W. (1979),1, Proceedings of the 6th Course of the International School of Cosmology and Gravitation on "Spin, Torsion and Supergravity" de ed. P.G. Bergamann  and V. de Sabatta , held at Erice .\\
{[7]} Hehl,F.W., McCrea, J.D. Mielke, E.W. and Ne'eman, Y.(1995)Phys Reports 1-171 \\
{[8]}Ali, S.A., Carfao, C. , Capozziello, S. and Corda, Ch. (2009) arXiv:0907.0934 \\
{[9]}Hayashi, K.  and Shirifuji, T. (1979) Phys. Rev. D, {\bf{19}}, 3524. \\
{[10]} Collins, P., Martin, A.  and Squires, E. "{\it{Particle Physics and Cosmology}}", John Wiley and Sons, New York.(1989) \\
{[11]}Hammond, R. (2002) Rep. Prog. Phys, 65, 599-449 \\
{[12]}Acros, H. I. and Pereira, J.G. (2004) International Journal of Modern Physics D  {\bf{13}}, 2193 \\
{[13]}Mao, Y, Tegmark, M., Guth, A. and Cabi, S. (2007)  Phys. Rev. D {\bf{76}}, 104029 ; arXiv:gr-qc/060812 \\
 {[14]} Hehl, F.W. (1971) Phys. Lett.36A, 225. \\
 {[15]} Hojman, S.(1978) Physical Rev. D  {\bf{18}}, 2741. \\
([16])Yasskin, P.H. and Stoeger, W.R. (1980) Phys. Rev. D 21, 2081 \\
 {[17]}Bazanski, S.L. (1989) J. Math. Phys., {\bf {30}}, 1018 ; \\ Kahil, M.E. (2006)  , J. Math. Physics {\bf {47}},052501. \\
{[18)} Kahil, M.E. (2015)   Odessa Astronomical Publications, {\bf{vol {28/2}}, 126.} \\
{[19]}  Mohseni, M. (2010), Gen. Rel. Grav., {\bf{42}}, 2477 . \\
{20} Acros, H. I., Andrade, V.C. and Pereira, J.G. (2004) arXiv;gr-qc/0403074 \\
{[21]} Hojman, S., Rosenbaum, M. and Ryan, M.P.(1979) Physical Rev. D {\bf{19}}, 430 \\
{[22]}Hehl, F.W., Obukhov, Yu, N. and Puetzfeld, D. (2013) A {\bf{377}}  1775; arXiv:1304.2769 \\
{[23]} Yisi, D. and Jiang, Y. (1999) Gen Rel. Grav. {\bf 31}, 99 . \\
{[24]} Cianfani, F., Montani, G. and Scopelliiti, V. (2015) arXiv: 1505.00943 \\
{[25]}Fabbri, L. and Vignolo, S. (2011) arXiv:1201.286 \\
{[26]} Wanas, M.I., Melek, M. and Kahil, M.E. (2000) Grav.
Cosmol., {\bf{6} }, 319. \\
{[27]} Wanas, M.I. and Kahil, M.E.(1999) Gen. Rel. Grav., {\bf{31}},
1921. ;\\Wanas, M.I., Melek, M. and Kahil, M.E. (2002) Proc. MG IX,
part B, p.1100, Eds.

V.G. Gurzadyan et al. (World Scientific Pub.); gr-qc/0306086.\\
{[28]} Wanas, M.I., Kahil, M.E. and Kamal, Mona. (2016) Grav.
Cosmol., {\bf{22} }, 345 . \\
{[29]}Puetzfeld,D. and Obukhov, Yu, N. (2013) AriXv:1308-2269 \\

\end{document}